\documentclass[aps,showpacs,showkeys,twocolumn]{revtex4}%
\usepackage{graphics,amsmath,amsfonts,amscd,revsymb,latexsym,
enumerate,multirow,epsfig}
\usepackage{amsmath}
\usepackage{amsfonts}
\usepackage{amssymb}
\usepackage{graphicx}%
\providecommand{\U}[1]{\protect\rule{.1in}{.1in}}
\providecommand{\U}[1]{\protect\rule{.1in}{.1in}}

\newcommand{\qed}{{\hfill$\Box$}}

\def\bi{\begin{itemize}}
\def\ei{\end{itemize}}
\def\be{\begin{equation}}
\def\ee{\end{equation}}
\def\bea{\begin{eqnarray}}
\def\eea{\end{eqnarray}}
\def\ben{\begin{eqnarray*}}
\def\een{\end{eqnarray*}}

\def\>{\rangle}
\def\<{\langle}

\newcommand{\1} I

\def\*{\star}

\def\0{{\mathbf{0}}}
\def\1{{\mathbf{1}}}
\def\2{{\mathbf{2}}}
\def\3{{\mathbf{3}}}
\def\4{{\mathbf{4}}}
\def\5{{\mathbf{5}}}
\def\6{{\mathbf{6}}}
\def\7{{\mathbf{7}}}
\def\8{{\mathbf{8}}}
\def\9{{\mathbf{9}}}

\begin{document}
\title{On the logical operators of quantum codes}
\author{Mark M. Wilde}
\email{mark.m.wilde@saic.com}
\affiliation{Electronic Systems Division, Science Applications International Corporation,
4001 North Fairfax Drive, Arlington, Virginia, USA\ 22203}
\date{\today }


\begin{abstract}
I show how applying a symplectic Gram-Schmidt orthogonalization to the
normalizer of a quantum code gives a different way of determining the code's
logical operators. This approach may be more natural in the setting where we
produce a quantum code from classical codes because the generator matrices of
the classical codes form the normalizer of the resulting quantum code. This
technique is particularly useful in determining the logical operators of an
entanglement-assisted code produced from two classical binary codes or from
one classical quaternary code. Finally, this approach gives additional
formulas for computing the amount of entanglement that an
entanglement-assisted code requires.

\end{abstract}
\maketitle

\section{Introduction}

The logical operators of a quantum code are important for fault-tolerant
quantum computation \cite{qecbook}\ and in the simulation of the performance
of stabilizer quantum error-correcting codes \cite{arx2007cross}. The typical
method for determining the logical operators of a quantum code is to employ
Gottesman's algorithm \cite{thesis97gottesman}. This algorithm brings the
stabilizer of a quantum code into a standard form and extracts the logical
operators from this standard form.

In this paper, I provide another technique for determining the logical
operators of a quantum code. This technique begins with the
\textit{normalizer} of the code and is perhaps more natural in the setting
where we produce a CSS\ quantum code from two classical additive binary codes
\cite{PhysRevA.54.1098,PhysRevLett.77.793}\ or where we produce a
CRSS\ quantum code from one classical additive code over GF$\left(  4\right)
$ \cite{ieee1998calderbank}. In both of these cases, we possess the full
normalizer of the resulting quantum code because the generator matrices of the
classical codes form its normalizer. Additionally, I show that this technique
is useful in determining the logical operators of an entanglement-assisted
code \cite{arx2006brun,science2006brun}, a type of quantum code to which
Gottesman's algorithm does not apply. The technique is again natural in the
setting where we import classical codes because the generator matrices form
the full normalizer of the resulting entanglement-assisted code. This
technique also gives another way of computing the amount of entanglement that
a given entanglement-assisted code requires (in this sense, this paper is a
sequel to the findings in Ref.~\cite{arx2008wildeOEA}). Specifically, I give a
set of formulas, different from those in Ref.~\cite{arx2008wildeOEA}, for
computing the entanglement that a given entanglement-assisted code requires.
The computation of the formulas in this paper may be more efficient than those
in Ref.~\cite{arx2008wildeOEA} and the efficiency depends on the parameters of
a given code. The technique for computing the logical operators and the
entanglement formulas in this paper should be a useful addition to the quantum
code designer's toolbox.

\section{Review of the Symplectic Gram-Schmidt Orthogonalization Procedure}

I first review the symplectic Gram-Schmidt orthogonalization procedure
(SGSOP)\ before proceeding to the main development of this paper. The SGSOP is
equivalent to finding a standard symplectic basis for a set of vectors
\cite{book2001symp}. It has made appearances in the quantum information
literature in entanglement-assisted quantum coding
\cite{arx2006brun,science2006brun},\ in finding an entanglement measure for a
stabilizer state \cite{arx2004fattal}, in the simulation of measurement
outcomes of stabilizer codes \cite{PhysRevA.70.052328}, and in the
decomposition of subsystem codes \cite{poulin:230504}.

We begin with a set of $m$ Pauli generators $g_{1}$, \ldots, $g_{m}$ that do
not necessarily form a commuting set. Consider generator $g_{1}$. There are
two possible cases for this generator:

\begin{enumerate}
\item Generator $g_{1}$ commutes with all other generators $g_{2}$, \ldots,
$g_{m}$. Set it aside in a \textquotedblleft set of
processed\ generators\textquotedblright, relabel generators $g_{2}$, \ldots,
$g_{m}$ as respective generators $g_{1}$, \ldots, $g_{m-1}$ and repeat the
SGSOP\ on these remaining generators.

\item Generator $g_{1}$ anticommutes with some other generator $g_{j}$ where
$2\leq j\leq m$. Relabel $g_{j}$ as $g_{2}$ and vice versa. Modify the other
generators as follows:%
\[
\forall i\in\left\{  3,\ldots,m\right\}  \ \ \ \ g_{i}=g_{i}\cdot
g_{1}^{f\left(  g_{i},g_{2}\right)  }\cdot g_{2}^{f\left(  g_{i},g_{1}\right)
},
\]
where the function $f$ is equal to zero if its two arguments commute and is
equal to one if its two arguments anticommute. (Note that we are not concerned
with the overall phases of any of the generators in this paper.) The new
generators $g_{3}$, \ldots, $g_{m}$ then commute with $g_{1}$ and $g_{2}$.
Place $g_{1}$ and $g_{2}$ into the \textquotedblleft set of processed
generators,\textquotedblright\ relabel generators $g_{3}$, \ldots, $g_{m}$ as
respective generators $g_{1}$, \ldots, $g_{m-2}$, and repeat the SGSOP\ on
these remaining generators.
\end{enumerate}

The result of the SGSOP\ is to produce a set of $m$ processed generators where
$2c$ of them form anticommuting pairs (each pair commuting with the generators
in the other pairs) and $l$ of them commute with themselves and the $c$ pairs
of generators. It is a reversible algorithm, meaning that we can recover the
original set of generators by performing all of the steps in reverse, and its
complexity is $O\left(  nm^{2}\right)  $. This complexity is the same as the
complexity of Gottesman's algorithm for bringing the generating set into a
standard form and determining the logical operators from this standard form
\cite{thesis97gottesman}.

\section{General Stabilizer Codes}

\label{sec:general-stab}We now consider applying the SGSOP\ to the normalizer
$N\left(  S\right)  $ of a stabilizer code \cite{thesis97gottesman}\ with
stabilizer $S$. The result is that the SGSOP\ gives a different way for
determining the logical operators of a given stabilizer code.

Suppose that we have a set $\left\langle S\right\rangle $ of $n-k$ Pauli
generators where each Pauli generator in $\left\langle S\right\rangle $ is an
$n$-fold tensor product of Pauli matrices. Suppose furthermore that
$\left\langle S\right\rangle $ generates an abelian group. Suppose we also
have a generating set $\left\langle N\left(  S\right)  \right\rangle $ of $p$
elements where $p>n-k$ and $\left\langle N\left(  S\right)  \right\rangle $
generates the normalizer $N\left(  S\right)  $ of the stabilizer $S$. The
relation $S\subseteq N\left(  S\right)  $ holds for this case because the set
of generators in $\left\langle S\right\rangle $ generates an abelian group.

We can perform the SGSOP\ on the generating set $\left\langle N\left(
S\right)  \right\rangle $. The SGSOP divides $\left\langle N\left(  S\right)
\right\rangle \,$into two different sets:\ the \textit{logical generating set
}$\left\langle S_{L}\right\rangle $ and the stabilizer generating set
$\left\langle S\right\rangle $. We can write the latter generating set as
$\left\langle S\right\rangle $ because it generates the same group that the
generating set from the previous paragraph generates. The logical generating
set $\left\langle S_{L}\right\rangle $ consists of $k$ anticommuting pairs of
generators that correspond to the logical operators for the stabilizer code
$S$. The size of the generating set $\left\langle N\left(  S\right)
\right\rangle $ is then $p=n-k+2k=n+k$.

We can also think about the code in terms of binary vectors that represent it
\cite{thesis97gottesman,book2000mikeandike}. Suppose we have the normalizer
matrix:%
\[
N\equiv\left[  \left.
\begin{array}
[c]{c}%
N_{Z}%
\end{array}
\right\vert
\begin{array}
[c]{c}%
N_{X}%
\end{array}
\right]  ,
\]
where $N$ is the binary representation of the generators in $\left\langle
N\left(  S\right)  \right\rangle $. We can consider the \textquotedblleft
symplectic product matrix\textquotedblright\ $\Omega_{N}$
\cite{arx2008wildeOEA} where%
\[
\Omega_{N}\equiv N_{Z}N_{X}^{T}+N_{X}N_{Z}^{T},
\]
and addition is binary. It is then straightforward to show by the method of
proof of Theorem~1\ in Ref.~\cite{arx2008wildeOEA}\ that%
\[
\frac{1}{2}\text{rank}\left(  \Omega_{N}\right)  =k.
\]
The idea of the proof is that the application of the SGSOP\ to the normalizer
matrix $N$\ transforms the matrix $\Omega_{N}$ to the following standard form:%
\begin{equation}%
{\displaystyle\bigoplus\limits_{i=1}^{k}}
J\oplus%
{\displaystyle\bigoplus\limits_{j=1}^{n-k}}
\left[  0\right]  ,
\end{equation}
where the small and large $\oplus$ correspond to the direct sum operation, $J$
is the matrix%
\begin{equation}
J=%
\begin{bmatrix}
0 & 1\\
1 & 0
\end{bmatrix}
,
\end{equation}
and $\left[  0\right]  $ is the one-element zero matrix. Each matrix $J$ in
the direct sum corresponds to a \textit{symplectic pair} and has rank two.
Each symplectic pair corresponds to exactly one logical operator in the code.
Each matrix $\left[  0\right]  $ has rank zero and corresponds to an ancilla
qubit. See Ref.~\cite{arx2008wildeOEA}\ for more details.

\subsection{CSS\ Codes}

One major class of quantum codes is the class of CSS\ codes
\cite{PhysRevA.54.1098,PhysRevLett.77.793}. These codes allow us to import two
\textquotedblleft dual-containing\textquotedblright\ classical codes for use
as a quantum code. The resulting quantum code inherits the error-correcting
properties and rate from the classical codes.

Here, we show that the SGSOP\ provides a way to determine the logical
operators for a CSS\ code. It constructs the full code (stabilizer and logical
operators) from the generator matrices of the two classical codes, rather than
exploiting the parity check matrices.

Suppose we have two classical codes $C_{1}$ and $C_{2}$ that satisfy
$C_{2}^{\perp}\subseteq C_{1}$. Suppose their respective generator matrices
are $G_{1}$ and $G_{2}$, and their respective parity check matrices are
$H_{1}$ and $H_{2}$. The conditions $H_{1}G_{1}^{T}=0$ and $H_{2}G_{2}^{T}=0$
follow from the definition of a parity check matrix, and the condition
$H_{1}H_{2}^{T}=0$ follows from the condition $C_{2}^{\perp}\subseteq C_{1}$.

The typical method for constructing a CSS\ quantum code is to build the
following stabilizer matrix:%
\[
\left[  \left.
\begin{array}
[c]{c}%
H_{1}\\
0
\end{array}
\right\vert
\begin{array}
[c]{c}%
0\\
H_{2}%
\end{array}
\right]
\]
The code forms a valid stabilizer matrix because $H_{1}H_{2}^{T}=0$ (this
condition corresponds to the requirement for a stabilizer code to consist of a
commuting set of generators).

The number of generators in the CSS\ quantum code is equal to $2n-k_{1}-k_{2}%
$, and it therefore has $n-\left(  2n-k_{1}-k_{2}\right)  =k_{1}+k_{2}-n$
logical qubits with $2\left(  k_{1}+k_{2}-n\right)  $ logical operators. The
size of the normalizer is thus $\left(  2n-k_{1}-k_{2}\right)  +2\left(
k_{1}+k_{2}-n\right)  =k_{1}+k_{2}$.

There is another way to build the code by exploiting the generator matrices
$G_{1}$ and $G_{2}$. The below matrix is the normalizer matrix of the code:%
\begin{equation}
\left[  \left.
\begin{array}
[c]{c}%
0\\
G_{2}%
\end{array}
\right\vert
\begin{array}
[c]{c}%
G_{1}\\
0
\end{array}
\right]  . \label{eq:CSS-normalizer}%
\end{equation}
The conditions $H_{1}G_{1}^{T}=0$ and $H_{2}G_{2}^{T}=0$ guarantee that the
above matrix is a valid normalizer matrix. It represents the full normalizer
for the code because the rows are linearly independent and the number of
generators in the normalizer is $k_{1}+k_{2}$.

Suppose that we run the SGSOP on the rows of the normalizer matrix
in\ (\ref{eq:CSS-normalizer}). The procedure then produces $k_{1}+k_{2}-n$
anticommuting pairs and $n-k_{1}+n-k_{2}$ commuting generators. The
$k_{1}+k_{2}-n$ anticommuting pairs correspond to $2\left(  k_{1}%
+k_{2}-n\right)  $ logical operators for the code, and the $n-k_{1}+n-k_{2}$
commuting generators correspond to the stabilizer generators of the code. This
technique for obtaining the logical operators and the stabilizer generators in
the case of a CSS\ code is perhaps more natural than applying Gottesman's
method in Ref.~\cite{thesis97gottesman} because we already know the matrices
$G_{1}$ and $G_{2}$ when importing classical codes.

By the same method of proof as Corollary~1 in Ref.~\cite{arx2008wildeOEA}, the
following formula also holds for any CSS\ code%
\begin{equation}
\text{rank}\left(  G_{1}G_{2}^{T}\right)  =k_{1}+k_{2}-n,
\label{eq:CSS-rank-comp}%
\end{equation}
because this quantity captures the amount of anticommutativity in the
generators in (\ref{eq:CSS-normalizer}).

\subsection{CRSS\ Quantum Codes}

A\ CRSS\ quantum code is one that we create by importing a classical additive
code over GF$\left(  4\right)  $ \cite{ieee1998calderbank}.

Suppose a GF$\left(  4\right)  $ code has parity check matrix $H$ and
generator matrix $G$ where $HG^{T}=0$. Additionally, suppose that the code is
dual under the trace product over GF$\left(  4\right)  $, i.e., the following
condition holds:%
\[
\text{tr}\left\{  HH^{\dag}\right\}  =0,
\]
where the dagger symbol $\dag$ is the conjugate transpose of matrices over
GF$\left(  4\right)  $ and the null matrix on the right has dimension $\left(
n-k\right)  \times\left(  n-k\right)  $. Note that the above trace operation
is different from the standard matrix trace, but is rather an elementwise
computation of the trace product over GF$\left(  4\right)  \ $%
\cite{arx2008wildeOEA}.

We can then create a quantum check matrix:%
\begin{equation}
\gamma\left(  \left[
\begin{array}
[c]{c}%
\omega H\\
\overline{\omega}H
\end{array}
\right]  \right)  , \label{eq:isomorphism-GF4-pauli}%
\end{equation}
where $\gamma$ is the isomorphism that takes an $n$-dimensional vector over
GF$\left(  4\right)  $ to a $n$-fold tensor product of Pauli operators
\cite{arx2008wildeOEA}.

It is straightforward to determine the normalizer of such a code because we
can construct it from the generator matrix $G$:%
\begin{equation}
\left[
\begin{array}
[c]{c}%
\omega G^{\ast}\\
\overline{\omega}G^{\ast}%
\end{array}
\right]  . \label{eq:gf4-normalizer}%
\end{equation}
The above matrix is a valid normalizer for the resulting quantum code because%
\begin{align}
\text{tr}\left\{  \left[
\begin{array}
[c]{c}%
\omega H\\
\overline{\omega}H
\end{array}
\right]  \left[
\begin{array}
[c]{c}%
\omega G^{\ast}\\
\overline{\omega}G^{\ast}%
\end{array}
\right]  ^{\dag}\right\}   &  =\text{tr}\left\{  \left[
\begin{array}
[c]{c}%
\omega H\\
\overline{\omega}H
\end{array}
\right]  \left[
\begin{array}
[c]{cc}%
\overline{\omega}G^{T} & \omega G^{T}%
\end{array}
\right]  \right\} \nonumber\\
&  =\text{tr}\left\{  \left[
\begin{array}
[c]{cc}%
HG^{T} & \overline{\omega}HG^{T}\\
\omega HG^{T} & HG^{T}%
\end{array}
\right]  \right\} \nonumber\\
&  =\left[
\begin{array}
[c]{cc}%
0 & 0\\
0 & 0
\end{array}
\right]  , \label{eq:reason-for-normalizer}%
\end{align}
where the first equality is by definition of the conjugate transpose, the
second equality holds because $\omega\overline{\omega}=1$, $\overline{\omega
}\overline{\omega}=\omega$, and $\omega\omega=\overline{\omega}$ for elements
of GF$\left(  4\right)  $. The last equality follows from the condition
$HG^{T}=0$.

We then can perform the SGSOP on the above normalizer matrix in
(\ref{eq:gf4-normalizer}). The result of the algorithm is to produce $2\left(
2k-n\right)  $ anticommuting pairs and $2\left(  n-k\right)  $ generators that
commute with themselves and the pairs. This method again determines the
logical operators for the imported code and may be a more natural alternative
to Gottesman's method because we already know the matrix $G$ that determines
the normalizer for the resulting CRSS\ quantum code.

By the same method of proof as Corollary~2 in Ref.~\cite{arx2008wildeOEA}, the
following formula holds for a trace-orthogonal code imported from GF$\left(
4\right)  $:%
\[
\text{rank}\left(  GG^{\dag}\right)  =2k-n.
\]

\section{Entanglement-Assisted Quantum Codes}

\label{sec:general-ent-stab}We can extend this procedure to an
entanglement-assisted code as well. This alternative procedure for determining
the logical operators is useful here because Gottesman's method does not apply
to a nonabelian group of Pauli generators.

Suppose that we have a set $\left\langle S\right\rangle $ of $m$ Pauli
generators that represents a generating set of operators with desirable
error-correcting properties. The generating set $\left\langle S\right\rangle $
does not necessarily generate an abelian group. Suppose we also have a
generating set $\left\langle N\left(  S\right)  \right\rangle $ of $p$
elements. This generating set $\left\langle N\left(  S\right)  \right\rangle $
generates the normalizer $N\left(  S\right)  $ of the stabilizer $S$. In
general, the relation $S\subseteq N\left(  S\right)  $ does not necessarily
hold for this case because the set of generators in $\left\langle
S\right\rangle $ is not an abelian group.

We can perform the SGSOP on the generators in $\left\langle S\right\rangle $.
This procedure divides $\left\langle S\right\rangle $ into two sets of
generators: the \textit{entanglement generator set} $\left\langle
S_{E}\right\rangle $\ and the \textit{isotropic generator set} $\left\langle
S_{I}\right\rangle $. The set $\left\langle S_{E}\right\rangle $ contains
anticommuting pairs of generators (where generators different pairs commute),
and the set $\left\langle S_{I}\right\rangle $ contains a commuting set of
generators that furthermore commute with the generators in $\left\langle
S_{E}\right\rangle $. The size of the generating set $\left\langle
S_{E}\right\rangle $ is $2c$, and the size of the set $\left\langle
S_{I}\right\rangle $ is $i$ so that $m=i+2c$.

We can also perform the SGSOP\ on the generating set $\left\langle N\left(
S\right)  \right\rangle $. The SGSOP\ divides this generating set into two
sets:\ the \textit{logical generating set} $\left\langle S_{L}\right\rangle $
and the isotropic generating set $\left\langle S_{I}\right\rangle $. We can
write this isotropic generating set as $\left\langle S_{I}\right\rangle $
because it generates the same group as the group generated by the isotropic
generating set from the previous paragraph. The size of the generating set
$\left\langle N\left(  S\right)  \right\rangle $ is then $p=i+2l$. The
generators in the logical generating set $\left\langle S_{L}\right\rangle $
are the logical operators for the code.

We can again consider this procedure in the binary vector representation.
Suppose the following matrix specifies the normalizer matrix:%
\[
N\equiv\left[  \left.
\begin{array}
[c]{c}%
N_{Z}%
\end{array}
\right\vert
\begin{array}
[c]{c}%
N_{X}%
\end{array}
\right]  .
\]
Consider the matrix $\Omega_{N}$ where%
\[
\Omega_{N}\equiv N_{Z}N_{X}^{T}+N_{X}N_{Z}^{T},
\]
and addition is binary. By the same method of proof as Theorem~1 in
Ref.~\cite{arx2008wildeOEA}, we can show that%
\[
\frac{1}{2}\text{rank}\left(  \Omega_{N}\right)  =l.
\]

\subsection{CSS\ Entanglement-Assisted Codes}

We now consider extending the development in
Section~\ref{sec:general-ent-stab}\ to the formulation of a CSS
entanglement-assisted code \cite{arx2006brun,science2006brun}. This extension
gives a useful way for determining the logical operators of a CSS
entanglement-assisted quantum code. Additionally, we determine a formula,
different from that in Corollary~1 of Ref.~\cite{arx2008wildeOEA}, for
computing the amount of entanglement that a given CSS entanglement-assisted
quantum code requires.

Consider two arbitrary additive binary codes $C_{1}$ and $C_{2}$ with
respective parity check matrices $H_{1}$ and $H_{2}$ and respective generator
matrices $G_{1}$ and $G_{2}$. We can form a quantum check matrix as follows:%
\begin{equation}
\left[  \left.
\begin{array}
[c]{c}%
H_{1}\\
0
\end{array}
\right\vert
\begin{array}
[c]{c}%
0\\
H_{2}%
\end{array}
\right]  . \label{eq:quantum-check-matrix}%
\end{equation}
The above matrix is not a valid stabilizer matrix, but we can run the
SGSOP\ on it to determine how to add entanglement in order to make this code a
valid stabilizer code. After adding ebits to the code, it then becomes an
entanglement-assisted stabilizer code. A CSS\ entanglement-assisted code
constructed in this way has $2n-k_{1}-k_{2}$ generators\ ($2c$ of which form
anticommuting pairs and $2n-k_{1}-k_{2}-2c$ form the isotropic generating
set), $k_{1}+k_{2}-n+c$ logical qubits, and consumes $c$ ebits
\cite{arx2006brun}.

The normalizer of the generators in (\ref{eq:quantum-check-matrix}) is again%
\[
\left[  \left.
\begin{array}
[c]{c}%
0\\
G_{2}%
\end{array}
\right\vert
\begin{array}
[c]{c}%
G_{1}\\
0
\end{array}
\right]  ,
\]
and represents the full normalizer of the generators in
(\ref{eq:quantum-check-matrix}) because the number of generators is
$k_{1}+k_{2}$ and the rows are linearly independent. This time, when we run
the algorithm, we get $2\left(  k_{1}+k_{2}-n+c\right)  $ anticommuting pairs
and $n-k_{1}+n-k_{2}-2c$ commuting generators. The commuting generators form
the basis for the isotropic generating set of the code and the anticommuting
generators correspond to the logical operators of the code.

By the method of proof of Corollary~1 in Ref.~\cite{arx2008wildeOEA}, we can
show that the following relation holds:%
\begin{equation}
\text{rank}\left(  G_{1}G_{2}^{T}\right)  =k_{1}+k_{2}-n+c.
\label{eq:EA-CSS-rank-comp}%
\end{equation}
This formula gives us another way to determine the amount of ebits that a
given entanglement-assisted code requires because the parameters $n$, $k_{1}$,
and $k_{2}$ are known parameters.

The complexity of the computation in (\ref{eq:EA-CSS-rank-comp}) is $O\left(
k_{1}k_{2}n\right)  $. A similar formula from Ref.~\cite{arx2008wildeOEA}
computes the amount of entanglement that the code requires:%
\[
\text{rank}\left(  H_{1}H_{2}^{T}\right)  =c.
\]
The complexity of the above formula is $O\left(  \left(  n-k_{1}\right)
\left(  n-k_{2}\right)  n\right)  $. Thus, the formula in
(\ref{eq:EA-CSS-rank-comp}) may provide a speedup for low-rate codes. It is up
to the quantum code designer to decide which formula to use depending on the
parameters of the code.

\subsection{CRSS\ Entanglement-Assisted Codes}

The method for constructing a CRSS\ entanglement-assisted code is again to
import an additive classical code over GF$\left(  4\right)  $. Suppose the
imported code has parity check matrix $H$ and generator matrix $G$ so that
$HG^{T}=0$. The imported code does not have to be orthogonal with respect to
the trace product. We then create a quantum check matrix:%
\[
\gamma\left(  \left[
\begin{array}
[c]{c}%
\omega H\\
\overline{\omega}H
\end{array}
\right]  \right)  ,
\]
where $\gamma$ is the same isomorphism as in (\ref{eq:isomorphism-GF4-pauli}).

The normalizer of the code is again the same as in (\ref{eq:gf4-normalizer}).
The normalizer matrix is%
\[
\gamma\left(  \left[
\begin{array}
[c]{c}%
\omega G^{\ast}\\
\overline{\omega}G^{\ast}%
\end{array}
\right]  \right)  ,
\]
and it is a valid normalizer for the same reasons as in
(\ref{eq:reason-for-normalizer}).

Performing the symplectic Gram-Schmidt orthogonalization procedure on the
normalizer matrix, we then get $2\left(  2k-n+c\right)  $ anticommuting pairs
and $2\left(  n-k-c\right)  $ commuting generators. This method again
determines the logical operators for the imported code.

By the same method of proof as Corollary~2 in Ref.~\cite{arx2008wildeOEA}, we
can show that the following formula holds:%
\begin{equation}
\text{rank}\left(  GG^{\dag}\right)  =2k-n+c. \label{eq:EA-gf4-formula}%
\end{equation}
The formula gives another way for determining the amount of entanglement that
an entanglement-assisted quantum code requires because the parameters $k$ and
$n$ are known parameters.

The complexity of the computation of rank$\left(  GG^{\dag}\right)  $ is
$O\left(  k^{2}n\right)  $. A similar formula from Ref.~\cite{arx2008wildeOEA}
computes the amount of entanglement that the codes requires:%
\[
\text{rank}\left(  HH^{\dag}\right)  =c.
\]
The complexity of the above formula is $O(\left(  n-k\right)  ^{2}n)$. Thus,
the formula in (\ref{eq:EA-gf4-formula}) may again provide a speedup for
low-rate codes.

\section{Conclusion}

I have shown how the normalizer of a quantum code, whether stabilizer or
entanglement-assisted, gives a useful way for determining its logical
operators. It is natural to assume full knowledge of the normalizer in the
case where we import classical codes to produce quantum codes. Additionally,
this development gives formulas for computing the amount of entanglement that
an entanglement-assisted code requires, and these formulas are an alternative
to those given in Ref.~\cite{arx2008wildeOEA}. It is straightforward to extend
these methods and formulas to qudit codes and continuous-variable codes, by
glancing at the results in Ref.~\cite{arx2008wildeOEA}.

Future work includes determining how to apply these findings in the setting of
quantum convolutional codes \cite{ieee2006grassl,ieee2007forney}\ and
entanglement-assisted quantum convolutional codes \cite{arx2007wildeEAQCC}.
Todd Brun and I have made attempts at this problem by devising the algorithm
in Section~IV-D\ of Ref.~\cite{arx2007wilde} for the CSS\ case and the
expansion technique and polynomial symplectic Gram-Schmidt algorithm in
respective Sections~IV and V of Ref.~\cite{arx2008wildeGEAQCC}\ for the
CRSS\ case. It is more complicated in the convolutional setting because we
would like to maintain the periodic structure of the quantum convolutional
code while still maintaining the standard symplectic relations. An algorithm
that achieves both tasks would be a boon to the theory of quantum
convolutional coding.

\begin{acknowledgments}
I thank Andrew Cross for useful comments on the manuscript and acknowledge the
support of an internal research and development grant SAIC-1669 of Science
Applications International Corporation.
\end{acknowledgments}

\bibliographystyle{unsrt}
\bibliography{Ref}

\begin{thebibliography}{10}

\bibitem{qecbook}
Frank Gaitan.
\newblock {\em Quantum Error Correction and Fault Tolerant Quantum Computing}.
\newblock CRC Press, Taylor and Francis Group, 2008.

\bibitem{arx2007cross}
Andrew~W. Cross, David~P. DiVincenzo, and Barbara~M. Terhal.
\newblock A comparative code study for quantum fault-tolerance.
\newblock {\em arXiv:0711.1556}, 2007.

\bibitem{thesis97gottesman}
Daniel Gottesman.
\newblock {\em Stabilizer Codes and Quantum Error Correction}.
\newblock PhD thesis, California Institute of Technology
  (arXiv:quant-ph/9705052), 1997.

\bibitem{PhysRevA.54.1098}
A.~Robert Calderbank and Peter~W. Shor.
\newblock Good quantum error-correcting codes exist.
\newblock {\em Physical Review A}, 54(2):1098--1105, Aug 1996.

\bibitem{PhysRevLett.77.793}
Andrew~M. Steane.
\newblock Error correcting codes in quantum theory.
\newblock {\em Physical Review Letters}, 77(5):793--797, Jul 1996.

\bibitem{ieee1998calderbank}
A.~Robert Calderbank, Eric~M. Rains, Peter~W. Shor, and N.~J.~A. Sloane.
\newblock Quantum error correction via codes over {GF(4)}.
\newblock {\em IEEE Transactions on Information Theory}, 44:1369--1387, 1998.

\bibitem{arx2006brun}
Todd~A. Brun, Igor Devetak, and Min-Hsiu Hsieh.
\newblock Catalytic quantum error correction.
\newblock {\em arXiv:quant-ph/0608027}, August 2006.

\bibitem{science2006brun}
Todd~A. Brun, Igor Devetak, and Min-Hsiu Hsieh.
\newblock Correcting quantum errors with entanglement.
\newblock {\em Science}, 314(5798):436--439, October 2006.

\bibitem{arx2008wildeOEA}
Mark~M. Wilde and Todd~A. Brun.
\newblock Optimal entanglement formulas for entanglement-assisted quantum
  coding.
\newblock {\em Physical Review A}, 77:064302, 2008.

\bibitem{book2001symp}
Ana~Cannas da~Silva.
\newblock {\em Lectures on Symplectic Geometry}.
\newblock Springer, 2001.

\bibitem{arx2004fattal}
David Fattal, Toby~S. Cubitt, Yoshihisa Yamamoto, Sergey Bravyi, and Isaac~L.
  Chuang.
\newblock Entanglement in the stabilizer formalism.
\newblock {\em arXiv:quant-ph/0406168}, 2004.

\bibitem{PhysRevA.70.052328}
Scott Aaronson and Daniel Gottesman.
\newblock Improved simulation of stabilizer circuits.
\newblock {\em Physical Review A}, 70(5):052328, November 2004.

\bibitem{poulin:230504}
David Poulin.
\newblock Stabilizer formalism for operator quantum error correction.
\newblock {\em Physical Review Letters}, 95(23):230504, 2005.

\bibitem{book2000mikeandike}
Michael~A. Nielsen and Isaac~L. Chuang.
\newblock {\em Quantum Computation and Quantum Information}.
\newblock Cambridge University Press, 2000.

\bibitem{ieee2006grassl}
Markus Grassl and Martin R\"{o}tteler.
\newblock Quantum convolutional codes: Encoders and structural properties.
\newblock In {\em Proceedings of the Forty-Fourth Annual Allerton Conference on
  Communication, Control, and Computing}, pages 510--519, 2006.

\bibitem{ieee2007forney}
G.~David Forney, Markus Grassl, and Saikat Guha.
\newblock Convolutional and tail-biting quantum error-correcting codes.
\newblock {\em IEEE Transactions on Information Theory}, 53:865--880, 2007.

\bibitem{arx2007wildeEAQCC}
Mark~M. Wilde and Todd~A. Brun.
\newblock Entanglement-assisted quantum convolutional coding.
\newblock {\em arXiv:0712.2223}, 2007.

\bibitem{arx2007wilde}
Mark~M. Wilde, Hari Krovi, and Todd~A. Brun.
\newblock Convolutional entanglement distillation.
\newblock {\em arXiv:0708.3699}, 2007.

\bibitem{arx2008wildeGEAQCC}
Mark~M. Wilde and Todd~A. Brun.
\newblock Quantum convolutional coding with shared entanglement: General
  structure.
\newblock {\em arXiv:0807.3803}, 2008.

\end{thebibliography}

\end{document}